\documentclass[aps,twocolumn,showpacs,superscriptaddress]{revtex4}
\usepackage{amsmath}
\usepackage{amsfonts} 
\usepackage{amssymb} 
\usepackage{graphicx} 

\newcommand{\td}{\tilde} 
\newcommand{\Tr}{\mathop{\mathrm{Tr}}}

\newcommand{\vecb}[1]{{\bf{#1}}}
\newcommand{\upd}{{\mathrm{d}}}

\newcommand{\nvechat}{\hat{\vecb n}}

\newcommand{\kvechat}{{\hat{\vecb k}}}

\newcommand{\zvechat}{\hat{\vecb z}}

\newcommand{\vvec}{\vecb v}

\newcommand{\Deltavec}{\boldsymbol{\Delta}}
\newcommand{\sigmavec}{\boldsymbol{\sigma}}

\newcommand{\omegaJ}{\omega_{J}}
\newcommand{\EJ}{E_{J}}
\newcommand{\FJ}{F_{J}}
\newcommand{\TJ}{T_{J}}
\newcommand{\Is}{I_{\mathrm{s}}}

\newcommand{\IS}{I^{\mathrm{S}}}
\newcommand{\IC}{I^{\mathrm{C}}}

\newcommand{\vF}{v_{\mathrm{F}}}
\newcommand{\vFvec}{\vvec_{\mathrm{F}}}

\newcommand{\Tc}{T_{\mathrm{c}}}

\newcommand{\kB}{k_{\mathrm{B}}}

\newcommand{\iu}{{\mathrm{i}}}

\newcommand{\im}{\mathop{\textrm{Im}}}
\newcommand{\re}{\mathop{\textrm{Re}}}

\newcommand{\lside}{\textrm{l}}
\newcommand{\rside}{\textrm{r}}

\newcommand{\spin}{\underline}

\begin{document}

\title{Dissipative Currents in
Superfluid $^3$He  Weak Links
}

\author{J. K. Viljas}
\affiliation{Low Temperature Laboratory, Helsinki University of
Technology, P.O.Box 2200, FIN-02015 HUT, Finland } 
\author{E. V. Thuneberg}
\affiliation{Low Temperature Laboratory, Helsinki University of
Technology, P.O.Box 2200, FIN-02015 HUT, Finland } 
\affiliation{Department of Physical Sciences,
P.O.Box 3000, FIN-90014 University of Oulu, Finland}

\date{\today}

\begin{abstract}
We calculate the current-pressure relation for pinholes connecting
two volumes of bulk superfluid $^3$He-B. The theory of
multiple Andreev reflections,  adapted from superconducting
weak links, leads to a nonlinear dependence of the dc current on
pressure bias. In arrays of pinholes one has to take into account 
oscillations of the texture at the Josephson frequency. The associated 
radiation of spin waves from the junction leads to an additional 
dissipative current at small biases, in agreement 
with measurements.

\end{abstract}

 
\maketitle

Weak links between superconducting metals have been extensively
studied during the past 40 years.
More recently, similar experiments have also been made in atomic
superfluids, in the fermion $^3$He liquid in particular \cite{josrev}. 
Experiments in superfluids can shed new light on general weak link
phenomena, due to the availability of parameter ranges which are not
easily realized in superconductors. There is also intrinsic interest
in superfluids since completely new phenomena may
appear which are not possible in conventional superconductors. 
Previous work on $^3$He weak links has mostly concentrated on determining the 
equilibrium properties, like the
Josephson current--phase relation $I(\phi)$. A new phenomenon identified
in this context is the {\em anisotextural} effect, where spin--orbit degrees
of freedom 
change as a function of the phase
difference $\phi$. As a consequence $I(\phi)$ deviates essentially
from the sinusoidal form
$I=I_c\sin\phi$ so that there appear
$\pi$ states, where both derivatives $I'(0)$ and $I'(\pi)$ are
positive
\cite{Viljas1,Viljas2}.   

In this letter we present a theory of dissipative currents in superfluid
$^3$He weak links.
Experimental studies of such time-averaged, or dc
currents under a 
pressure bias have been reported in Refs.\
\onlinecite{Steinhauer} and \onlinecite{Simmonds}. At constant chemical
potential difference $U$ 
the supercurrent $I(\phi)$ oscillates at the Josephson frequency
$\omega_J=2U/\hbar$. Therefore the superfluid part does not
contribute to the dc current unless there is a
mechanism for absorbing the extra energy $2U$ per transported
Cooper pair.
One example is the process of multiple Andreev reflections (MAR),
used for explaining properties of superconducting weak links 
\cite{gz,Averin}. Due to Andreev reflection there are
quasiparticle states bound to the contact region. These states
carry the current, one Cooper pair per one back-and-forth reflection.
Simultaneously the quasiparticles gain energy
until they escape above the gap 
or are relaxed by scattering. 
This process works also in superfluid
$^3$He, but with important differences in details.
Below we argue that in $^3$He there is an additional dissipation mechanism.
Owing to the anisotextural effect, energy can be carried away from the junction
by spin-wave radiation. We show that the
two mechanisms together can  explain the essential features of the
experimental results in Refs.\ \onlinecite{Steinhauer} and
\onlinecite{Simmonds}.

\emph{Multiple Andreev reflections.---}Let us consider a weak link
under constant difference $U$ in chemical potentials. In
$^3$He $U$ is related to the pressure difference $P$ by
$U=(m_3/\rho)P$, where $\rho$ is the density and $m_3$ the mass of
an atom. The time derivative of the phase difference $\phi$ across the
link is given by the Josephson frequency
$\dot\phi=\omega_J=2U/\hbar$. The time-dependent
mass current can then be expressed as a Fourier series  
\begin{equation} \label{e.fullc}
\begin{split}
I(P,t)=I_{0}(P)+\sum_{n=1}^\infty[
\IS_{n}(P)\sin(n\omegaJ t) 
+
\IC_{n}(P)\cos(n\omegaJ t)
]
\end{split}
\end{equation}
with coefficients $I_0,\IS_n,\IC_n$. We have calculated these
coefficients for a pinhole, where the dimensions of the
hole are assumed small in comparison with the superfluid coherence
length 
$\xi_0= 77$ nm. (All parameters are evaluated at vapor pressure.)
Our calculation uses nonequilibrium Green's functions, and is
explained in the appendix. However, a large part of the results can be
understood by the following adiabatic model, which is a 
generalization of the one used for superconductors \cite{Averin}.

The current arises from quasiparticle states that are bound in the weak
link at energies below the gap $\Delta$. 
The mass current is given by
\begin{equation}
I(t)=\frac{m_3}{\hbar}\sum_i\sum_\delta\sum_n\sum_\sigma
\frac{d\epsilon_{i\delta n\sigma}}{d\phi}p_{i\delta n\sigma}.
\label{e.isum}\end{equation}
Here $\epsilon_{i\delta n\sigma}(\phi)$ is the bound state energy that depends
on the channel index $i$, momentum sign $\delta=\pm$, and spin index
$\sigma=\pm$. The index
$n$ is for the case that there exist more than one bound state. The
distribution function
$p_{i\delta n\sigma}(t,\phi)$ obeys the kinetic equation
\begin{equation}
\frac{\partial p_{i\delta n\sigma}}{\partial t}
+\dot\phi
\frac{\partial
p_{i\delta n\sigma}}{\partial \phi}=\Gamma[f(\epsilon_{i\delta
n\sigma})-p_{i\delta n\sigma}],
\label{e.kinetic}\end{equation}
where $f(\epsilon)=[\exp(\epsilon/\kB T)+1]^{-1}$ is the
Fermi distribution and $\Gamma^{-1}$ a
relaxation time.

\begin{figure}[!tb]
\begin{center}
\includegraphics[width=0.7\linewidth]{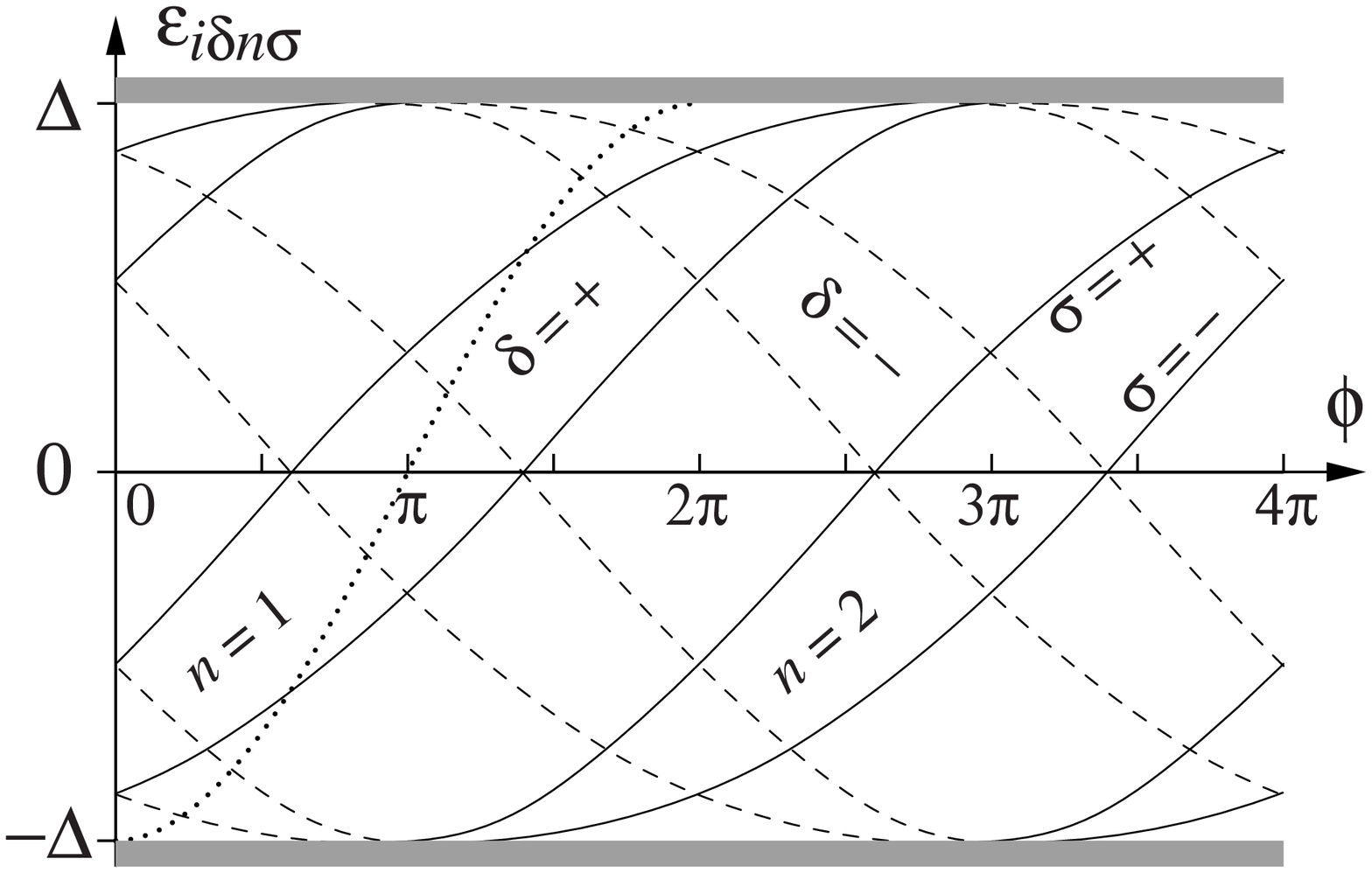}
\caption{The bound state energies $\epsilon_{i\delta n\sigma}$ as a function
of the phase difference $\phi$. Neglecting gap suppression
$\epsilon_\delta=-\delta\Delta\cos(\phi/2)$ (dotted line). Parameters
are: quasiparticle direction cosine 0.93, temperature $0.6
T_c$ and 
$\textsf{R}^{\lside,\rside}=\textsf{R}(\mp\zvechat,\theta_L)$.}
\label{f.disp}
\end{center}
\end{figure}

Application of the above to $^3$He includes a few specific features. 
First, the order parameter is always suppressed near walls. We calculate
it self-consistently, assuming diffusive scattering of quasiparticles 
at the wall. This implies that also the energies
$\epsilon_{i\delta n\sigma}(\phi)$ have to be
calculated numerically. One example is shown in Fig.\ \ref{f.disp}.
Second, since we consider the B phase, the bulk order parameter is 
of the form 
$\Delta \textsf{R}\exp(\iu\phi)$, where
$\textsf{R}$ is a rotation matrix \cite{vw}. Depending on the matrices
on the left ($\lside$) and right ($\rside$)
hand sides, $\textsf{R}^{\lside,\rside}$, there is
spin-splitting:  
$\epsilon_{i\delta n-}(\phi)\approx\epsilon_{i\delta n+}(\phi-\psi)$
\cite{Viljas4,Yip}. Third, the only important  source of inelastic
processes is  quasiparticle-quasiparticle collisions. Near a surface
the order parameter is suppressed and therefore we approximate the
scattering rate by the normal Fermi-liquid form
$\Gamma=a[(\pi \kB T)^2+\epsilon^2]/(\tau_0\pi^2\kB^2)$. Here 
$\tau_0=1.14\,\mu{\rm s}\,{\rm mK}^2$ is obtained from viscosity
measurements and
$a$ is a coefficient on the order of unity. 
The low-energy $\Gamma$ is denoted by $\Gamma_0=aT^2/\tau_0$.

\begin{figure}[!tb]
\begin{center}
\includegraphics[width=0.9\linewidth]{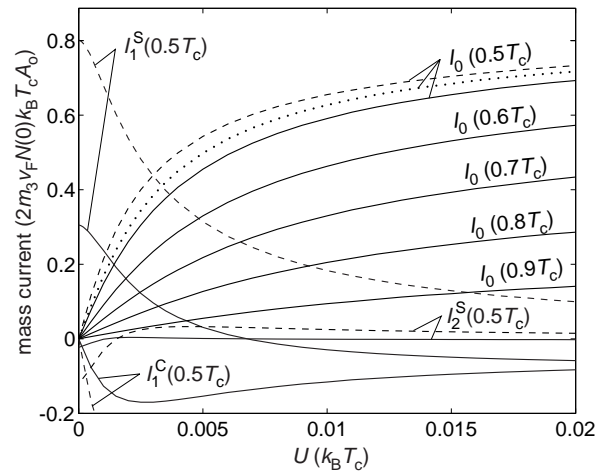}
\caption{Current amplitudes 
 $I_0$, $\IS_1$, $\IS_2$ and $\IC_1$ (\ref{e.fullc}) as functions of
the chemical potential difference $U$. The average current
$I_0$ is shown at different temperatures, other curves are for
$T/\Tc=0.5$. $I_0$ is linear at small $U$ (\ref{e.arc}) and
saturates at larger $U$, as discussed in the main text. The most
accurate results are shown by solid lines. The effect of neglecting the
gap suppression is shown by dashed lines, and the effect of neglecting
the energy dependence of
$\Gamma$ is shown by a 
dotted line. 
The largest shown
bias is $P\approx 4.2$ Pa. Parameters are
$a=1.6$ and
$\textsf{R}^\lside=\textsf{R}^\rside$. }
\label{f.fig1}
\end{center}
\end{figure}
We are now equipped to analyze Fig.\ \ref{f.fig1}.
Similar to the experiments, we study the low-bias region,
$U\ll k_BT_c$. In this region we can neglect the normal current from
energies above $\Delta$,
excluding only a narrow temperature slice near the superfluid
transition temperature
$T_c$. A characteristic scale for $U$ is set by the scattering strength
$\hbar\Gamma$. At $U=0$, the distribution function is in equilibrium, 
$p_{i\delta n\sigma}=f(\epsilon_{i\delta n\sigma})$. Equation
(\ref{e.isum}) then gives the equilibrium current-phase relation
$I(\phi)$, dominated by $\sin(\phi)$ (at least for
$\textsf{R}^\lside=\textsf{R}^\rside$) but with smaller admixtures of
$\sin(n\phi)$ where $n>1$. 
At $U\ll\hbar\Gamma$ the kinetic equation (\ref{e.kinetic}) can
easily be solved by linearization. This leads to a time-independent
component
$I_{0}$ that is linear in $U$ and independent of 
$\textsf{R}^{\lside,\rside}$,
\begin{equation} \label{e.arc}
I_{0}=g(T)({\Delta}/{\hbar\Gamma_0})G_{\textrm{n}}
U.
\end{equation}
Here $G_{\textrm{n}}$ is the 
normal-state conductance and $g(T)$ a factor on the order of
unity. Neglecting gap suppression,  $g(T)=\int_{-1}^1\tanh(\Delta
x/2\kB T)(x/\sqrt{1-x^2})\upd x$.
The validity region of the linear dependence (\ref{e.arc}) vanishes as
$T\rightarrow 0$ and is replaced by
$I_{0}\propto U^{1/3}$.  Finally, at
$U\gg\hbar\Gamma$ (but still
$U\ll\Delta$) the scattering can be neglected and $p$ is determined by
thermalization at gap edges, $p_{i\delta n\sigma}=f(-\delta\Delta)$.
Here $I_{0}$ approaches a constant value that is on the same order as
the critical current. For a self-consistent order parameter all
higher harmonics are effectively damped by the smooth
$\epsilon_{i\delta n\sigma}(\phi)$.

The $I_0(P)$ in Fig.\ \ref{f.fig1} can be compared with the
measurements in Ref.\ \onlinecite{Steinhauer}. They have
the same shape and a good order-of-magnitude
agreement on both axes. The agreement is surprisingly good taking
into account that the aperture sizes are on the order of $\xi_0$ or
larger, and thus our pinhole approximation is not justified.

The more recent experiment of Ref.\ \onlinecite{Simmonds} was done
with an array of apertures. In contrast to Ref.\
\onlinecite{Steinhauer}, this experiment should be well in the region
where the linear approximation (\ref{e.arc}) is valid.  However, a
clearly nonlinear
$I-P$ curve was measured.  The experimental results are also different
for the two possible states of the weak link, the ``H'' and the ``L''
states, which have previously  been identified as two nearly degenerate
textural states with different $\textsf{R}^{\lside,\rside}$'s
\cite{Viljas2}. 
Yet, our numerical calculations confirm that 
$I_0(P)$ curves in Fig.\ \ref{f.fig1} remain practically unchanged
for all $\textsf{R}^{\lside,\rside}$. This can also be seen in the adiabatic
model as follows. The spin splitting
$\epsilon_{i\delta n-}(\phi)\approx\epsilon_{i\delta n+}(\phi-\psi)$ has
an essential effect on the critical current so that $\IS_1$ can
even vanish \cite{Yip}. 
In $I_0$, however, the relative
phase shift is unimportant since the phase runs through all values.   

We conclude that
the nonlinearity and the texture-dependence in
Ref.\ \onlinecite{Simmonds} are either large-aperture effects, and/or
result form something other than  MAR. In the following we demonstrate
that at least part of the H-L difference can be accounted for by the
anisotextural effect, which exists in array-type weak links of
$^3$He-B.

\emph{The anisotextural model.---}As discussed above, the
current--phase relation $I(\phi)$ depends on the rotation matrices
$\textsf{R}^{\lside,\rside}$. The basic idea of the anisotextural
effect is that  
$\textsf{R}^{\lside,\rside}$ are not fixed, but  tend to move toward their
$\phi$-dependent equilibrium configuration. This effect was previously
used to explain the so-called $\pi$ states, where a local minimum of
energy appears at
$\phi\approx\pi$ \cite{Viljas2}.
Below we generalize this theory to the dynamical case
where $\phi(t)=\omegaJ t$.

\begin{figure}[!tb]
\begin{center}
\includegraphics[width=0.8\linewidth]{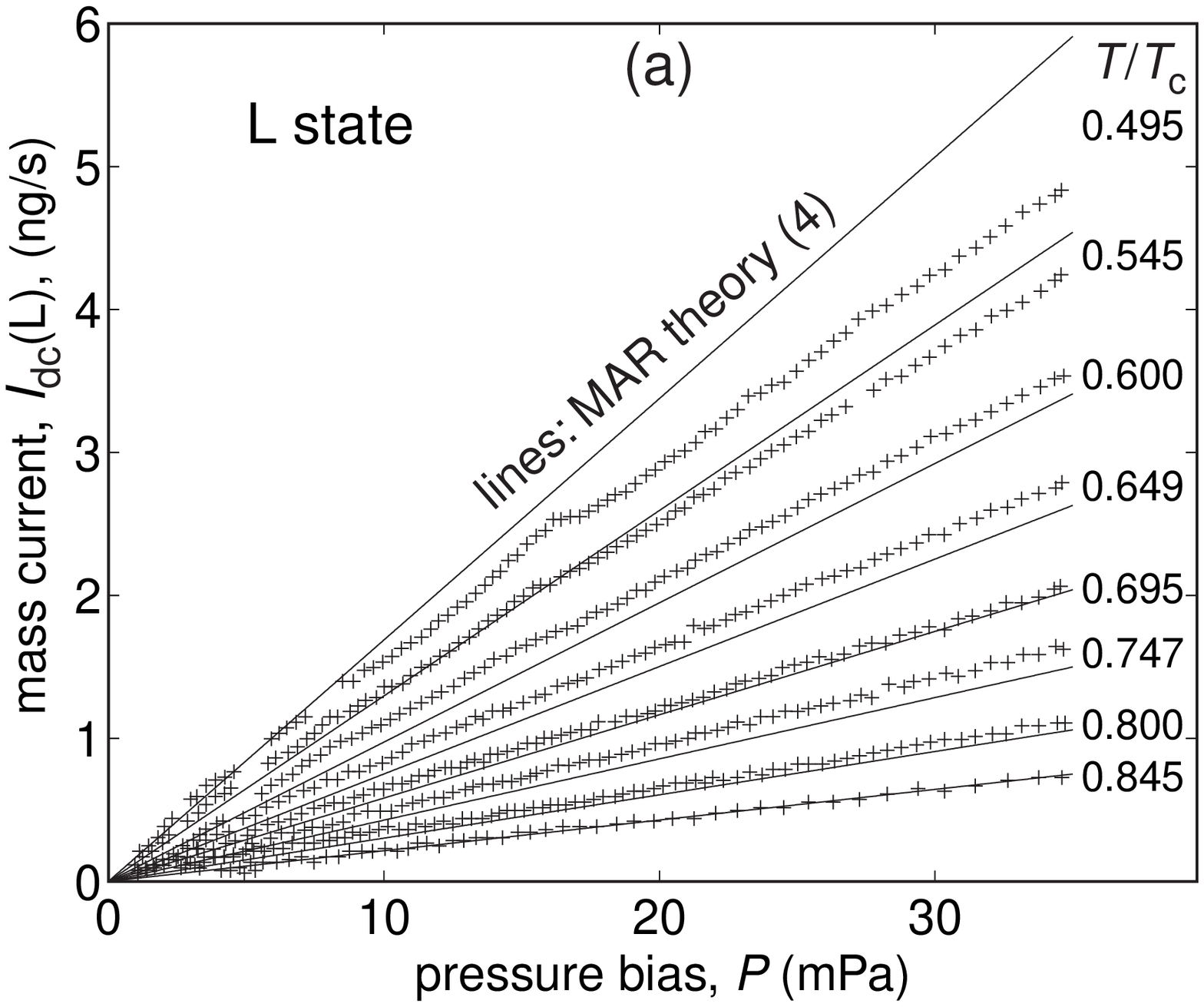}
\includegraphics[width=0.8\linewidth]{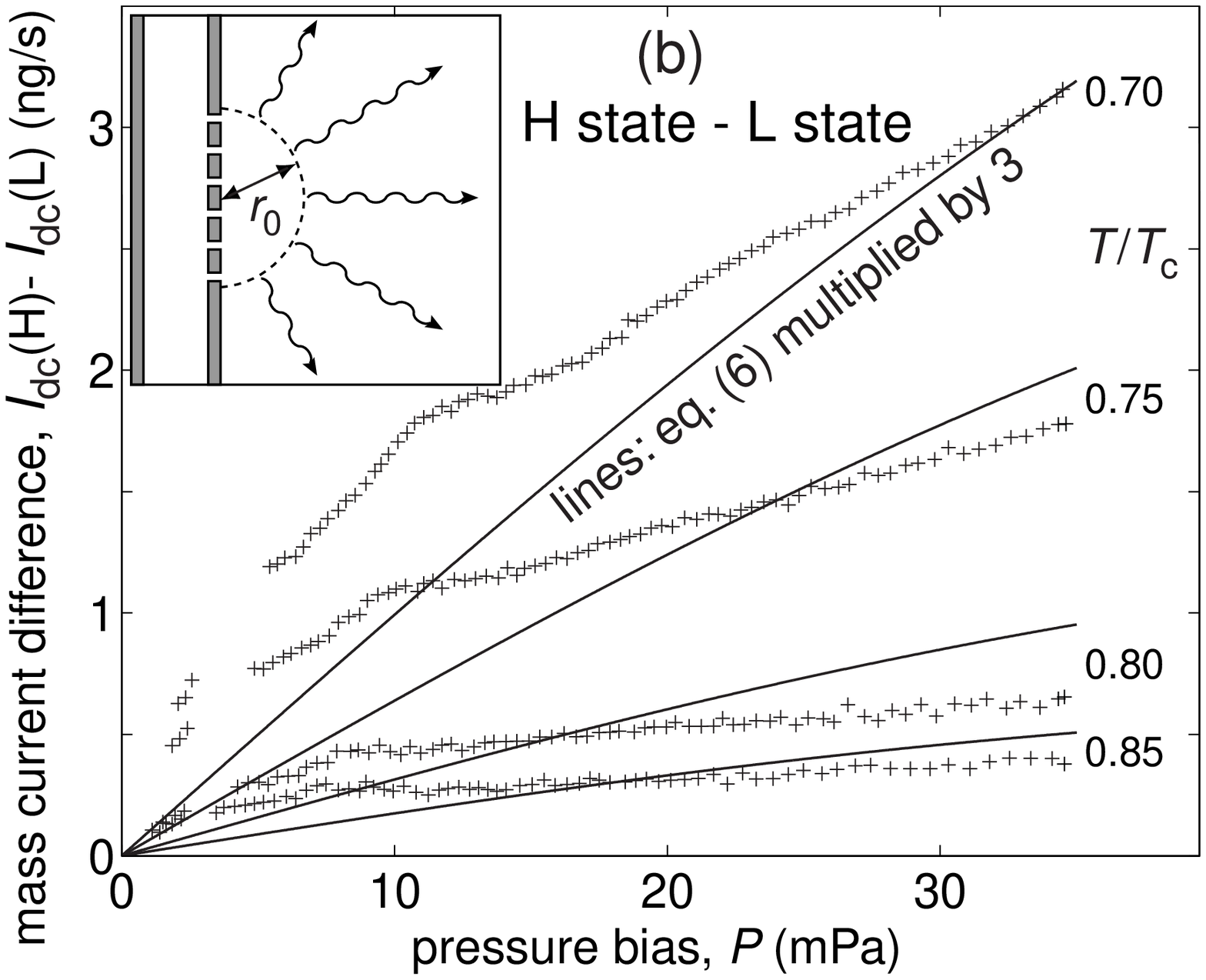}
\caption{Comparison of experimental \cite{Simmonds} ($+$)
and theoretical (solid lines) current-pressure relations. (a) The L
state current is compared with theoretical $I_0$ (\ref{e.arc}) with a
single adjustable parameter $a=1.6$ coming from the collision rate
$\Gamma$. (Alternatively, one can neglect the gap suppression and use
$a=3.2$.) 
(b) The difference between the H and L state currents is compared with
the difference of
$I_{\textrm{ai}}$ (\ref{e.isave}), which is arbitrarily multiplied by
factor 3. The inset
depicts the Josephson array, cut-off radius $r_0$, and radiation of
spin waves.}
\label{f.fig2}
\end{center}
\end{figure}

A simple but still realistic model for the anisotextural effect in 
an array of apertures is
based on the energy functional \cite{Viljas2}
\begin{equation} \label{e.energy}
F[\eta] = \FJ(\eta_0,\phi)
+\textstyle{\frac{1}{2}}K\int\upd^3r |\nabla\eta|^2.
\end{equation}
Here $\FJ$ is the Josephson
coupling energy and the second term is
due to the bending of the rotation axis $\nvechat$ which
parametrizes the matrix $\textsf{R}(\nvechat,\theta_L)$.
The rotation angle is fixed to
$\theta_L\approx 104^\circ$ by the bulk dipole-dipole 
interaction \cite{Leggett}. 
Due to the geometry of the experiment \cite{Simmonds}, 
we assume $\nvechat$ to be fixed 
parallel to the wall normal $\zvechat$
on one side of the array.
On the other side $\nvechat$ makes an angle $\eta(r)$ with 
$\zvechat$, which depends on
the distance $r$ from the center of the array.
We choose a cutoff at $r=r_0$, below which
$\eta(r)$ is assumed constant (see inset in Fig.\ \ref{f.fig2}).
The $F_J$ term depends on the angle
$\eta_0\equiv\eta(r_0)$ near the
junction, and the bending term is nonzero if 
$\eta_0$ differs from $\eta_\infty\equiv\eta(\infty)$ preferred 
by walls and other textural interactions.
At  temperatures near
$T_c$ the energy $F_J$ can be approximated by
$\FJ=-\EJ\cos\phi$ where 
$\EJ=\alpha R_{\mu z}^\lside R_{\mu z}^\rside+$
$\beta(R_{\mu x}^\lside R_{\mu x}^\rside+R_{\mu y}^\lside R_{\mu y}^\rside)$.
We choose to fix the
 $\rside$ side 
($\nvechat^\rside\parallel\zvechat$),
and on the $\lside$ side it is convenient to take
$0<\eta_\infty<\pi/2$. The H and L states are then identified as 
the states with
$\nvechat^\rside=\pm\zvechat$, respectively \cite{Viljas2}. 
Linearizing around
$\eta\approx\eta_\infty$,  we find 
$\EJ \approx \EJ^\infty-J_{\rm sp}\vartheta(r_0)$, where 
$\vartheta=\eta-\eta_\infty$,  and
$J_{\rm sp}(\eta_\infty)=[\pm(20\alpha+5\beta)\sin 2\eta_\infty$
$+30\beta\sin\eta_\infty]/16$.
The coefficients $\alpha$, $\beta$ and 
$K$ are calculated in Ref.\ \onlinecite{Viljas2}.

The dynamics is determined by the Leggett equations 
$\dot S=-\delta F/\delta\theta$ and $\dot
\theta=\mu_0\gamma_{\textrm{g}}^2S/\chi$ \cite{Leggett}. Here the
spin $S$ and spin-rotation angle $\theta = \sqrt{5/2}\vartheta$
are conjugate variables.
$\gamma_{\textrm{g}}$ is the gyromagnetic ratio and $\chi$ the spin
susceptibility. [The Leggett equations are canonical equations
for a Hamiltonian where the spin energy density
$\mu_0\gamma_{\textrm{g}}^2S^2/2\chi$ is added to $F$ in 
Eq.\ (\ref{e.energy})]. The dynamical equations reduce to a wave
equation in the bulk and $\FJ$ determines a boundary
condition at the junction. The solution at constant
$P$ consists of waves that propagate radially out of the
junction. This implies that the supercurrent 
$\Is=(2m_3/\hbar)\partial_{\phi}\FJ(\eta_0,\phi)$
has a nonzero average, the anisotextural current
\begin{equation} \label{e.isave}
I_{\textrm{ai}}(P)=\frac{2m_3}{\hbar}
\frac{[J_{\rm sp}(\eta_\infty)]^2}{4\gamma}
\frac{\omega_J r_0/c}{1+(\omega_J r_0/c)^2}.
\end{equation}
Here $c=\sqrt{2\mu_0\gamma_{\rm g}^2 K/(5\chi)}$ is the spin-wave
velocity and $\gamma=b\pi K r_0$. We estimate 
$r_0=\sqrt{A/\pi}$, where 
$A$ is the surface area of the array. We also take into account
that the actual geometry around the junction differs from a half-space
and use $b=0.31$ instead of unity \cite{Viljas2}. 

The crucial features of the anisotextural current $I_{\textrm{ai}}$
(\ref{e.isave}) are the dependence on the textural configuration via
$J_{\rm sp}(\eta_\infty)$, and the nonlinear dependence on pressure via
$\omega_J$. The origin of $I_{\textrm{ai}}$ is the
oscillating part in Eq.\ (\ref{e.fullc}), which was here approximated
by the $\IS_{1}$ term alone. The total dc current
$I_{\textrm{dc}}$ is the sum of $I_{\textrm{ai}}$
and the constant component due to MAR, 
$I_{\textrm{dc}}=I_{0}+I_{\textrm{ai}}$. 

\emph{Comparison to experiments.---}
It turns out that the L state 
current, which is rather linear \cite{Simmonds}, can almost
completely be attributed to MAR. 
The comparison of the L state data with $I_0$
is presented in Fig.\ \ref{f.fig2}(a). 
There is only one
adjustable parameter $a\sim 1$, which is associated with the uncertainty
of the relaxation rate $\Gamma$. 
A likely reason for the differences 
is that the
theory uses the pinhole approximation,
which is not strictly valid for the experimental apertures (100
nm$\times$100 nm squares in a wall of thickness 50 nm are not small
compared to $\xi_0= 77$ nm).  

Since both the H and L state data are assumed to contain the same
contribution $I_0$, it is convenient to subtract the H and L state
data taken at equal temperatures. In this way one may directly compare
the texture-dependent parts with $I_{\rm ai}$, regardless of any
uncertainty that may be present in $I_0$.  This is done in  Fig.\
\ref{f.fig2}(b). The theoretical result corresponds to the difference
between Eq.\ (\ref{e.isave}) calculated for the two states. There are
no adjustable parameters, since the only free parameter is the
textural angle
$\eta_\infty$, which was previously found to be approximately 
$0.3\pi$ based on static properties \cite{Viljas2}. The anisotextural
model, as presented above, can explain roughly one third of the
observed H-L difference. [The theoretical current in Fig.\
\ref{f.fig2}(b) is arbitrarily multiplied by factor 3 in order to
make it better visible.] Also, the curvature in the theoretical lines
is at higher biases than in the experimental results. There are several
possible sources for the differences,  in particular the
oversimplifications used in our anisotextural model. Unfortunately, it
would be very demanding to improve upon the pinhole approximation, or
to calculate the texture and propagation of spin waves in the
complicated geometry of the experiment \cite{Viljas2}.  There is also
uncertainty in the experimental 
parameters, for example in the diameter of the apertures, which
appears in its fourth power in 
$I_{\textrm{ai}}$. As a result, the
true reason for the factor-of-three discrepancy remains open. 

\emph{Conclusion.---}We have presented a theory of dissipative
currents in $^3$He weak links. It shows that the nonlinearities in
the measurements of Refs.\
\onlinecite{Steinhauer} and \onlinecite{Simmonds} have different
origins, and both can semi-quantitatively be explained by natural
extensions of existing theories. The extension of static anisotextural
phenomena to dynamics gives further support for the theory, and
provides the energy loss mechanism that is required for a weak
link to become trapped in a $\pi$ state, as seen in experiments
\cite{josrev}. The anisotextural phenomena are sensitive to the
experimental cell and magnetic field, for example, and thus can be
tested in detail in future experiments. Also the oscillating components
$\IS_n$ and $\IC_n$ in Fig.
\ref{f.disp} as well as the spin waves might be observable in
experiments.  

\emph{Appendix: Calculation of currents.---}Consider a 
pinhole  with open area $A_o$. Assume the $\lside$-side
chemical potential to be shifted by 
$U$ with respect to the $\rside$ side, and take
the $z$ axis to point from $\lside$ to $\rside$. 
We define $G_{\textrm{n}}$ in Eq.\ (\ref{e.arc}) as 
$G_{\textrm{n}}=\frac{1}{2}m_3\vF N(0)A_o$, where 
$N(0)$ is the single-spin density of states in the normal state 
and $\vF$ the Fermi velocity
\cite{Viljas2}. The mass current may then be written as
$I(t)=G_{\textrm{n}}\langle \hat k_z I(\kvechat,t)\rangle_{\hat k_z>0}$
where $\langle\cdots\rangle_{\hat k_z>0}$
$=\int_{\hat k_z > 0}(\upd\Omega_\kvechat/4\pi)\cdots$
denotes an average over the Fermi-surface points $\kvechat$.
Since $I(t)$ is periodic with period $\TJ=2\pi/\omegaJ$, we expand
$I(\kvechat,t)=\sum_{n=-\infty}^{\infty}
I_n(\kvechat)e^{\iu n\omegaJ t}$
such that
$I_n(\kvechat)=I^*_{-n}(\kvechat)$.
Using the $\spin\gamma$ matrices of Ref.\ \onlinecite{Eschrig} to
expand the Keldysh function, the amplitudes for $n\geqslant 0$ may 
be written as 
$I_n(\kvechat)=\Tr\spin C \{ 
2U\delta_{n0} 
-\sum_{m=0}^\infty 
\mathcal P\int\upd\epsilon 
[\spin F_{\lside\rside}^{m,m+n}(\kvechat,\epsilon,U)- 
\spin F_{\rside\lside}^{m+n,m}(-\kvechat,\epsilon,-U)]\}$
where
$\spin F_{ij}^{l,m}(\kvechat,\epsilon,U)=
\spin P^m_{ij}(\epsilon,U)
[\spin x_{}^i(\epsilon)-
\spin\gamma_{}^{Ri}(\epsilon)\td{\spin x}_{}^j(\epsilon-U)
\td{\spin\gamma}_{}^{Ai}(\epsilon)]
[\spin P^l_{ij}(\epsilon,U)]^\dagger$
and
$\spin P^m_{ij}(\epsilon,U) = 
\prod_{p=0}^{m-1}
\spin\gamma^{Ri}_{}(\epsilon+(2p+2)U)
\td{\spin\gamma}^{Rj}_{}(\epsilon+(2p+1)U)$
where $i,j=\lside,\rside$, $\spin C=\spin 1$, and 
$\kvechat$ dependences are omitted for clarity.
(For spin current, replace $m_3\rightarrow\hbar/2$ 
and $\spin C\rightarrow\spin\sigmavec$.)
Since we are considering a pinhole, the $\lside$ and $\rside$ 
order parameters
may be assumed to be unaffected by the weak link and to be in equilibrium. 
Thus the distribution functions are of the form
$\spin x^i_{}(\epsilon)=h^i(\epsilon)$
$[\spin 1-\spin\gamma^{Ri}_{}(\epsilon)$
$\td{\spin\gamma}^{Ai}_{}(\epsilon)]$ and
$h^i(\epsilon)=1-2f(\epsilon)$.
The conjugation symbol ``$~\td{~}~$'' is defined as
in Ref.\ \cite{Eschrig}
and the amplitudes satisfy
$\td{\spin\gamma}^A=[\spin\gamma^R]^\dagger$.

Above the $\spin\gamma^{R,A}$'s refer to the coherence 
functions inside the junction.
They are obtained by integrating the Riccati
equation 
$\iu\hbar\vFvec\cdot\nabla\spin\gamma^R_0 = -2\epsilon^R\spin\gamma^R_0$
$-\spin\gamma^R_0\spin\Delta_0^\dagger\spin\gamma^R_0-\spin\Delta_0$
on several trajectories toward the wall on both sides, 
where $\epsilon^R=\epsilon+\iu(\hbar\Gamma/2)$,
$\spin\Delta_0=\Deltavec_0\cdot\spin\sigmavec\iu\spin\sigma_2$, 
$\Deltavec_0(\kvechat,z)=$
$(\Delta_\parallel\hat k_x,\Delta_\parallel\hat k_y,$
$\Delta_\perp\hat k_z)$, 
and $\Delta_\parallel(z)$, $\Delta_\perp(z)$ are calculated with the
``ROM'' model of a diffusive surface (cf. Ref.\ \cite{Viljas2}).
For a symmetrical junction, the $\lside$ and $\rside$ solutions 
at the junction satisfy 
$\spin\gamma_0^{R\rside}(-\kvechat)=-[\spin\gamma_0^{R\lside}(\kvechat)]^T$.
The different spin rotations 
$\spin U^{\lside,\rside}$
$=\exp(-\iu\theta_L\nvechat^{\lside,\rside}\cdot\spin\sigmavec/2)$
are taken into account by writing 
$\spin\gamma^{Ri}=\spin U^i\spin\gamma_0^{Ri}[\spin U^i]^T$,
for $i=\lside,\rside$. 

The current amplitudes of Eq.\ (\ref{e.fullc}) in Fig.\ \ref{f.fig1}
are obtained with 
$I_0=G_{\textrm{n}}\langle \hat k_zI_0(\kvechat)\rangle_{\hat k_z>0}$,
$\IS_n=2G_{\textrm{n}}\im\langle \hat k_zI_n(\kvechat)\rangle_{\hat k_z>0}$,
$\IC_n=2G_{\textrm{n}}\re\langle \hat k_zI_n(\kvechat)\rangle_{\hat k_z>0}$.
These results have been derived for time-independent
$\textsf{R}^{\lside,\rside}$, but
at least for $U\ll\hbar\Gamma_0$ we may expect them to be 
valid even if the textures vary on time scale $T_J$.



\begin{thebibliography}{9}

\bibitem{josrev} J.C. Davis and R.E. Packard, Rev. Mod. Phys. {\bf 74},
741 (2002).

\bibitem{Viljas1} J. K. Viljas and E. V. Thuneberg, 
Phys. Rev. Lett. \textbf{83}, 3868 (1999).

\bibitem{Viljas2} J. K. Viljas and E. V. Thuneberg, 
Phys. Rev. B \textbf{65}, 064530 (2002).

\bibitem{Steinhauer} J. Steinhauer, \emph{et al.},
Physica B \textbf{194-196}, 767 (1994).

\bibitem{Simmonds} R. W. Simmonds, \emph{et al.},
Phys. Rev. Lett. \textbf{84}, 6062 (2000).

\bibitem{gz} U. Gunsenheimer and A. D. Zaikin, 
Phys. Rev. B \textbf{50}, 6317 (1994).

\bibitem{Averin} D. Averin and A. Bardas, Phys. Rev. Lett. \textbf{75},
1831 (1995);
Phys. Rev. B \textbf{53}, R1705 (1996).

\bibitem{vw}
D. Vollhardt and P. W\"olfle, 
\emph{The superfluid phases of helium three} 
(Taylor \& Francis, London, 1990).

\bibitem{Viljas4} J. K. Viljas and E. V. Thuneberg,
J. Low Temp. Phys. \textbf{134}, 743 (2004).

\bibitem{Yip} S.-K. Yip, 
Phys. Rev. Lett \textbf{83}, 3864 (1999).

\bibitem{Leggett} A.J. Leggett, Ann. Phys. (New York) \textbf{85}, 11
(1974).

\bibitem{Eschrig} M. Eschrig, 
Phys. Rev. B \textbf{61}, 9061 (2000).





\end{thebibliography}
\end{document}